%% file: main.tex
\documentclass[11pt,a4paper,superscriptaddress]{revtex4-2}
\usepackage{amsthm}
\usepackage{graphicx}  
\usepackage{dcolumn}   
\usepackage{bm}        
\usepackage{amssymb}   
\usepackage{amsmath}
\usepackage{epsfig}
\usepackage{rotating}
\usepackage{xspace}
\usepackage{hyperref}
\usepackage{xcolor}
\usepackage{soul}

\newcommand{\C}{\mathbb{C}}

\newcommand{\R}{\mathbb{R}}

\renewcommand{\Re}{{\rm Re}}
\renewcommand{\Im}{{\rm Im}}

\newcommand{\cH}{{\cal H}}
\newcommand{\cL}{{\cal L}}

\newcommand{\cO}{{\cal O}}

\newcommand{\bra}{\langle}
\newcommand{\ket}{\rangle}

\newcommand{\bear}{\begin{eqnarray}}
\newcommand{\eear}{\end{eqnarray}}

\def\be#1\ee{\begin{equation}#1\end{equation}}
\def\bea#1\eea{\begin{align}#1\end{align}}

\begin{document}

\title
{Complex Langevin: Correctness criteria, boundary terms and spectrum}

\author{Erhard Seiler}
\email{ehs@mpp.mpg.de}
\affiliation{Max-Planck-Institut f\"ur Physik (Werner-Heisenberg-Institut),
F\"ohringer Ring 6, M{\"u}nchen, Germany}
\author{D\'enes Sexty}
\email{denes.sexty@uni-graz.at}
\affiliation{
Institute of Physics, NAWI Graz, University of Graz, Universitätsplatz 5, Graz, Austria}
\author{Ion-Olimpiu Stamatescu}
\email{I.O.Stamatescu@thphys.uni-heidelberg.de}
\affiliation{ Institut f\"ur Theoretische Physik, Universit\"at Heidelberg, Heidelberg, Germany}

\date{\today}

\begin{abstract} 

The Complex Langevin (CL) method to simulate `complex probabilities', ideally produces 
expectation values for the observables that converge to a limit equal to the 
expectation values obtained with the original complex `probability' measure. The 
situation may be spoiled in two ways: failure to converge and convergence to the wrong 
limit. It was found long ago that `wrong convergence' is caused by boundary terms; 
non-convergence may arise from bad spectral properties of the various evolution 
operators related to the CL process. Here we propose a class of criteria which allow to rule out boundary terms and at the same time bad spectrum. Ruling out boundary terms in the equilibrium distribution arising from a CL simulation implies that the so-called convergence conditions are fulfilled. This in turn has been shown to guarantee that the expectation values of holomorphic observables are given by complex linear combinations of $\exp(-S)$ over various integration cycles. If the spectrum is pathological, however, the CL simulation in general does not reproduce the integral over the desired real cycle.

\end{abstract}
\maketitle

\input 1intro.tex
\input 2general.tex

\input 3asalcedo.tex

\input 3bsalcedo.tex
\input 4xymodel.tex
\input 5discussion_rev.tex
\input 6appendixa.tex

\input 7appendixb.tex

\input 8references.tex
\end{document}

%% file: 1intro.tex
\section{Introduction}

The Complex Langevin (CL) stochastic process \cite{parisi,klauder} describes the evolution of a probability distribution on the complexified configuration space, generated by the 
Fokker-Planck (FP) operator $L^T$. The probability distribution is supposed to converge 
to an equilibrium distribution which reproduces the averages with respect to a complex 
measure of a class of holomorphic observables. 
In the last two decades the Complex Langevin method enjoyed renewed interest,
it has been tested as a proposed solution to the sign problem in various systems
such as the real time evolution of quantum field theories
\cite{Berges:2005yt,Berges:2006xc,Berges:2007nr,alvestad,Boguslavski:2022dee,Lampl:2023xpb,Alvestad:2023jgl}, various 
systems having nonzero chemical potential \cite{Aarts:2008wh,Aarts:2011zn,Mollgaard:2014mga,Bloch:2017sex} and
condensed matter systems \cite{Hayata:2014kra,Berger:2019odf}. 
The introduction of gauge cooling \cite{Seiler:2012wz} made simulations possible also in gauge 
theories, allowing complex Langevin simulations in QCD to be carried out \cite{Sexty:2013ica,Nagata:2018mkb,Kogut:2019qmi,Sexty:2019vqx,Scherzer:2020kiu,Ito:2020mys,Attanasio:2022mjd}.

The application of the Complex Langevin equation is not without problems, though.
Two possibilities of failure have been identified: the process may fail to converge and it
may converge to the wrong limit. Failure to converge may occur via exponentially
increasing expectation values of certain observables, related to spectrum in the right half complex plane, or it may occur
due to spreading of the probability measure, creating slowly decaying `skirts'. In the latter case expectation
values of observables with high powers become mathematically undefined (thus fail to converge in practice).

The conservation of probability under the CL process guarantees a certain trivial 
stability: expectation values of observables which are bounded on the complexified 
configuration space will remain bounded under the CL process. This shows that the 
semigroup generated by FP operator is bounded on an appropriate (Banach) space and 
exponentially growing modes are not present for bounded observables. But unfortunately 
this is irrelevant, as we need to consider holomorphic observables, which are not 
bounded unless they are constant. So for the study of convergence and stability a 
different mathematical setting is needed. This is done in the next section.

More serious is the problem of wrong convergence. This is typically due to the 
spreading out of the probability density, leading to slow decay (skirts) of the 
distribution and the occurrence of boundary terms, as discussed in 
\cite{boundaryterms,boundaryterms2}. This spreading can be tested by the control variables introduced in 
the next section. These are used to define a general class of criteria for correctness 
of the CL results; they are designed to rule out boundary terms but they also rule out 
instability in the form of exponential or sub-exponential growth in the evolution of observables. 
The criteria are sufficient, provided certain additional conditions hold; so they are 
not eliminating {\em all} possible failures of the CL method. These limitations as well 
as open problems are discussed in the last section.
  
We check some versions of the criteria numerically for a simple model of one variable 
in Section 3 and for a lattice model in Section 4. By direct numerical determination 
of the spectrum it is revealed that violation of the criteria does not necessarily 
imply the presence of `bad' spectrum. There is a regime in which failure of the 
criteria indicate presence of boundary terms and `wrong convergence' of the CL 
simulations takes place, yet the the direct determination of the spectrum shows absence 
of exponentially growing modes. In a simple model we find that the appearance of 
unwanted spectrum is linked to the Lee-Yang zeroes; the significance of this fact is not 
yet entirely clear.
 
A word of caution is in order: the spectrum of a formally defined differential operator 
depends on the precise definition of the space on which it operates; likewise the 
relation between the spectrum of an operator and the behavior of the semigroup it 
generates may be more subtle than we are used to from finite dimensions. Some of these 
points are addressed in the appendices.

There are, in principle, four possible combinations of boundary terms or no boundary terms, and bad spectrum or no bad spectrum (by bad spectrum we mean eigenvalues of $L_c$, the complex Fokker-Planck operator, with positive real part). We find that, depending on the parameters chosen, bad spectrum and boundary terms may appear together, but boundary terms may also appear without bad spectrum and, of course, there also is also a regime in which neither boundary terms nor bad spectrum occur. 
So in the presence of boundary terms there is always either wrong convergence or no
convergence. Hence the absence of boundary terms is a crucial condition for
correctness; the absence of bad spectrum
plays only a subsidiary role. Limitations of this statement, concerning situations where
the absence of boundary terms, while necessary, is not sufficient for correctness, are
discussed in Section \ref{limitations}.

%% file: 2general.tex
\section{Mathematical generalities}

To keep the notation simple, we carry out this discussion for one variable. The generalization 
to many variables requires some care, as discussed in Section \ref{xy}.

\subsection{Notation}

We consider complex measures given by a complex density $\rho$ which is holomorphic and
given in terms of an action as
\be
\rho(z)=\exp(-S(z))
\ee
Expectation values of holomorphic observables observables $\cO$ are given by integration
over a suitable integration ``cycle'' (in the terminology of Witten \cite{witten_ams})
$\gamma$:
\be
\bra \cO(z)\ket= \frac{1}{Z}\int_\gamma \rho(z)\cO(z) dz\,;\quad Z=\int_\gamma \rho(z)dz\,.
\ee
The CL equation is
\be
dz(t) = K(z) dt + dw(t)\,;\quad  K(z)=\rho'(z)/\rho(z)=-S'(z)\,; 
K_x=\Re\, K\,,K_y=\Im\, K\,,
\ee
where $dw$ is the increment of the Wiener process normalized as
\be
\bra dw(t)^2 \ket =2 dt\,.
\ee
The evolution of the probability density $P$ on $\C=\R^2$ is given by The Fokker-Planck(FP)
equation:
\be
\partial_t P(x,y;t)=L^T P(x,y;t)\,,\quad L^T=\partial_x^2-\partial_x K_x-\partial_y K_y\,.
\ee
The evolution of observables is given by transpose $L$ of $L^T$
\be
\partial_t \cO(x,y;t)=L \cO(x,y;t)\,,\quad L=\partial_x^2+K_x\partial_x+K_y \partial_y\,,
\ee
which simplifies for holomorphic observables to
\be
\partial_t \cO(z;t)=L_c \cO(z;t)\,,\quad L_c =\partial_z^2 +K(z) \partial_z\,.
\ee
Formally, these linear evolutions are solved by exponential semigroups, such as $\exp(tL)$ 
etc..

\subsection{A trivial fact}

We start with a simple fact. Our probability measures on $\C\equiv \R^2$ are given by 
distributional `densities' $P(x,y;t)$ on $\R^2$, i.~e. positive distributions (this 
includes $\delta$ distributions and $\cL^1$ functions). We denote the standard (total 
variation) norm of measures on $\R^2$ by $||.||_1$. For a probability measure $P$ we have
\be
||P||_1=1\,.
\ee
Since $\exp(tL^T)$ preserves probability, it is a contraction on the space of complex 
measures, i.~e. for any complex density $\rho$ and all $t\ge 0$,
\be
||\exp(tL^T)\rho||_1\le ||\rho||_1\,.
\ee
This means in particular that as an operator on $\cL^1(\R^2)$, $L^T$ has no unstable
(exponentially growing) modes. More explicitly this can be seen by noting that the
Fokker-Planck evolution operator $\exp(tL^T)$ has an integral kernel
$\exp(tL^T)(x,y;x',y')\ge 0$ satisfying
\be
\int dx'dy'\exp(tL^T)(x,y;x',y')=1\,.
\ee
Mathematically the natural space of observables would be the space of bounded 
continuous functions, a subspace of  $\cL^\infty$ in $\C$. On this space $\exp(tL)$ is really the adjoint (transpose) of $\exp(tL^T)$ and
\be
\int P(x,y;0) (\exp(tL)\cO)(x,y) dx dy\le ||\cO||_\infty\equiv \sup_{x,y}|\cO(x,y)|\,,
\ee
so the dual semigroup $\exp(tL)\cO$ is again a contraction (i.e. has $\infty$ norm $\le 1$).

Unfortunately $\cL^\infty(\R^2)$ does not contain any nonconstant holomorphic 
functions, so clearly the space of observables has to be enlarged; that requires the space of allowed measures to be restricted in such a way that all observables in the enlarged space have well-defined expectation values.

\subsection{Conditions on the space of probability measures}

To get a nontrivial space of holomorphic observables, we have to consider weighted 
spaces of measures and observables. We define these spaces in terms of a strictly 
positive weight function $\sigma(x,y)$, growing at infinity, using the norms
\be
||f||_{1,\sigma}=\int \sigma(x,y) |f(x,y)| dx\,dy, \quad
||\cO||_{\infty,\sigma^{-1}}=\sup_{x,y} |\cO(x,y)/\sigma(x,y)|\,.
\ee
By a slight abuse of notation we denote the space of complex measures $\rho$ on $\C$ 
with $||\rho||_{1,\sigma}<\infty$ by $\cL^1_\sigma$ and the space of observables $\cO$ 
with $||\cO||_{\infty,\sigma^{-1}}<\infty$ by $\cL^\infty_{1/\sigma}$. Of course the 
space of observables also has to be chosen in such a way that the evolution operator 
$L_c$ and the semigroup $\exp(tL_c)$ leave it invariant; this means that we cannot 
limit ourselves to a finite dimensional subspace, such as polynomials up to a given 
order.

The choice of the weight $\sigma$ is dictated by the class of observables we want to
consider: it has to be chosen such that they lie in $\cL^\infty_{\sigma^{-1}}$. In other
words: $\sigma$ has to grow at least as fast as all the observables we want to consider.
We should also choose $\sigma$ so that it does not grow too much faster than the observables 
in our chosen space, because we do not want it to cause a `false alarm' about slow decay.

For the noncompact case of $\R^d$ the usual set of observables consists of all polynomials,
so we need $\sigma$ to grow faster than any power at infinity. Possible choices are
\be
\sigma_1(x,y)=\exp(\alpha(x^2+y^2)^{1/4})\,,\quad \alpha>0\,
\label{sigma1}
\ee
or
\be
\sigma_2(x,y)=\exp(\alpha(\log(1+x^2+y^2))^2)\,,\quad \alpha>0\,.
\label{sigma2}
\ee
In the following we use the second choice $\sigma\equiv \sigma_2$ with $\alpha=1$.

For the case of a compact real configuration space, such as $U(1)$, the natural space 
of observables is spanned by the exponentials $\exp(inz)$, which grow exponentially in 
the imaginary direction. So in this case we need a stronger than exponential decay in 
the measure; we may choose for instance
\be
\sigma_3(x,y)=\exp(|y|^\alpha)\,,\;\alpha>1\,,
\label{sigma3}
\ee
or
\be
\sigma_4(x,y)=\exp\left(\alpha|y|\log(1+|y|)\right)\,,\alpha>0\,.
\label{sigma4}
\ee

A similar growth of $\sigma$ is needed for other compact groups like $SU(N)$ and hence
for lattice gauge models.

It is not automatically true that $\exp(tL^T)$ is a contraction from $\cL^1_\sigma$ to 
itself; we have to make an assumption quantifying the necessary decay of the measure 
evolving under the CL process:

{\em Assumption $A$:}
\be
||\exp(tL^T) P||_{1,\sigma}\le C ||P||_{1,\sigma}\equiv C_\sigma\;\;  
\forall P\in \cL^1_\sigma,\; t\ge 0\,,
\label{assumption}
\ee
with a constant $C_\sigma$ independent of $t$. Written out, (\ref{assumption}) says
\be
\int P(x,y;t)\sigma(x,y) dx\,dy \le C_\sigma\,.
\label{assumption_equiv}
\ee
The weight function plays the role of a {\em non-holomorphic} control variable, which is 
required to have a well-defined and bounded expectation value under the probability measure
$P$ evolving according to the Fokker-Planck equation.  

If the CL process is ergodic, the choice of the initial distribution $P(x,y;0)$ does not 
matter and we may for instance choose $P(x,y;0)=\delta(x)\delta(y)$ which is
convenient for numerical checks.

Eq. (\ref{assumption}) or (\ref{assumption_equiv} imply
\be
\left|\int P(x,y;t) \cO(x,y) dx\,dy\right|\le C_\sigma||\cO||_{\infty,\sigma^{-1}}\,,
\ee
so expectation values of observables will be bounded in time; i.~e. Assumption A implies 
that there are no exponentially growing modes showing up.

Assumption $A$ is a criterion for correctness: Since it guarantees strong enough decay 
on $P(x,y;t)$, it implies the absence of boundary terms for the observables in 
$\cL^\infty_{1/\sigma}$.

Condition $A$ has the following relation to the `drift criterion' of \cite{nagata}: the 
latter can be interpreted as a special case; it requires the existence of an $\alpha>0$ 
such that Assumption $A$ is satified for the choice $\sigma=\sigma_{d,\alpha}$ with
\be
\sigma_{d,\alpha}(x,y)\equiv \exp(\alpha|K(x+iy)|)\,.
\label{sigmad}
\ee
This shows that for any polynomial $S$ of higher than second degree, the drift 
criterion is stronger than the versions (\ref{sigma1}) and (\ref{sigma2}). In compact 
cases, the drift grows exponentially in the noncompact directions, so the control 
variable $\sigma_{d,\alpha}$ is again stronger than (\ref{sigma3}) and (\ref{sigma4}). It 
is in fact stronger than necessary, i.~e. it might signal incorrectness of certain CL 
results when they are in fact correct. This has been found to actually occur in some 
cases of the one-link $U(1)$ model \cite{unpub} but it might also happen for polynomial 
models.

Failure of Assumption $A$ indicates insufficient deacy of the probability distribution,
i.~e. the presence of skirts. It should not be confused with `runaways', i.~e. breakdown of the simulation after a finite time; this problem was eliminated in all cases encountered by the use of adaptive step size \cite{Aarts:2009dg}. 

\subsection{Implications for the spectrum}

The Langevin operator $L$ is the formal transpose of $L^T$. {\em It is the true transpose if
and only if there are no boundary terms.} Since Assumption $A$ guarantees the absence of
boundary terms for observables in the appropriate space $\cL^{\infty}_{1/\sigma}$, we have
indeed
\be
\int P(x,y;t) \cO(x,y) dx\,dy = \int P(x,y;0) \cO(x,y;t)dx\,dy\,,
\label{nobound}
\ee
where $\cO(x,y;t)=(\exp(tL)\cO)(x,y)$ and $P(x,y;t)=(\exp(tL^T)P)(x,y)$.

Under Assumption $A$ the left hand side of (\ref{nobound} )is bounded uniformly in $t$, 
so the right hand side is uniformly bounded as well i.~e. there are no unstable modes 
of $L$ showing up.

In more detail the argument goes as follows: by assumption
\be
\left|\int P(x,y;t) \cO(x,y) dx\,dy\right|\le C ||P||_{1,\sigma} ||\cO(x,y)||_{\infty,1/\sigma}\,;
\label{bound1}
\ee
choosing now
\be
P(x,y;0)=\delta(x-x_0)\delta(y-y_0)
\ee
this bounds the left hand side of (\ref{bound1}) by
\be
C\sigma(x_0,y_0) ||\cO(x,y;0)||_{\infty,1/\sigma}\,.
\ee
If there is no boundary term, i.~e. (\ref{nobound}) holds, we thus find
\be
|\cO(x_0,y_0;t)|\le C\sigma(x_0,y_0) ||\cO(x,y;0)||_{\infty,1/\sigma}\,
\label{stabil}
\ee
which shows the absence of unlimited growth (exponentially or otherwise) in $\cO(x+iy;t)$.

Restricting $L$ to the subspace of holomorphic functions in $\cL^\infty_{\sigma^{-1}}$,
it can be replaced by $L_c$ due to the Cauchy-Riemann equations, so under our assumption,
$L_c$ as well can have no unstable modes in this space.

%% file: 3asalcedo.tex
\section{Quartic model}

L.~L.~Salcedo \cite{salcedo} pointed out that a simple quartic model sheds light on some
problems of the CL method. The model is defined by the action
\be
S(x)=\frac{\lambda}{4}x^4+ \frac{m^2}{2}x^2+hx,
\label{salcedo}
\ee
which we investigate for $\lambda\ge 0$, $m^2>0$ and complex $h$ (below we denote the imaginary part of $h$ with $h_I$), corresponding to the complex density
\be
\rho(x)\equiv \exp(-S(x))\,.
\ee
(Note that a lattice version of this model was studied in \cite{Attanasio:2021tio} using the Complex Langevin equation.) 
As remarked by Salcedo, given $\lambda>0, m^2\ge 0$, the partition function $Z(h)=\int\rho(x) 
dx$ vanishes for certain values purely imaginary of $h$ (so-called Lee-Yang zeroes), leading to divergent 
expectation values of $x^n$, whereas the CL equation does not show anything special at these 
values; so clearly the CL results cannot be correct. This fact is borne out by numerical 
studies, which show deviations between the exact results and the numbers produced by CL, 
becoming most dramatic near the Lee-Yang zeroes. Here we want to point out that these deviations 
are linked to massive failures of assumption $A$ for $\Im\ h$ larger than some value $h_c$, so Eq. 
(\ref{stabil}) does not hold there, we expect boundary terms to occur and we cannot use the criterion 
to rule out spectrum of $L_c$ in the right half plane. In the next subsection the spectrum of 
$L_c$ is directly determined numerically with the result that the appearance of unstable modes, while not coinciding with the appearance of boundary terms, 
is actually linked to the Lee-Yang zeroes (see also \cite{Attanasio:2021tio}).

To give a definite meaning to the spectrum of $L_c$, we consider it as an operator in the 
Hilbert space
\be
\cH_S=\cL^2(\R,\exp(-\Re S)dx)\,.
\label{hilbertdef}
\ee
The spectrum is then the same as the spectrum of
\bea
-H=& \exp(-S/2) L \exp(S/2)\notag \\
=&\partial_x^2+\frac{m^2}{2}+\frac{h^2}{4}+(\frac{3\lambda}{2}+\frac{m^4}{4})x^2
-\frac{\lambda m^2}{2}x^4-\frac{\lambda^2}{4}x^6+\frac{h m^2}{2}x+\frac{\lambda h}{2}x^3\,.
\eea
considered as an operator on 
\be
\cH=\cL^2(\R,dx). 
\ee
(Maybe it would be more natural to continue working in the Banach space defined in the previous
section, but for spectral considerations Hilbert spaces are more convenient.)

We are interested in complex `magnetic fields' $h$, so $H$ is a Schr\"odinger operator 
with a complex potential.

Let's now consider purely imaginary $h$: 
\be
h=ih_I\,,\quad h_I\in \R\,; 
\ee
the spectrum of hermitian part $(H+H^\dagger)/2$ now reaches down to $-h_I^2/4<0$; in fact
\be
\psi^{(0)}_0(x)=\exp(-\Re~S(x)/2)\,,
\ee 
is an eigenvector of $(H+H^\dagger)/2$ with eigenvalue $-h^2/4$. This shows that
$\exp(tL_c)$ is {\bf not} a contractive semigroup on $\cH_S$. But the numerics presented in the
next subsection suggests that nevertheless
\be
||\exp(tL_c)||_{\cH_S} =||\exp(-tH)||_\cH
\ee
remains bounded for all $t>0$, provided $h_I$ is small enough. $\psi_0=\exp(-S(x)/2)$ is an 
eigenvector of $-H$ with eigenvalue $0$ and presumably $\exp(tL_c)\phi$ converges to a multiple 
of $\psi_0$ for all $\phi\in \cH_S$ in this case (see discussion in Appendix A.)

\subsection{Checking Assumption $A$}

We choose the weight function
\be
\sigma_2(x,y)\equiv \exp[(\log(1+x^2+y^2))^2]\,; 
\label{sigma0}
\ee
this allows to consider observables in the (Banach) space defined by the norm
\be
||\cO(x+iy)||_{\infty,1/\sigma}=\sup_{x,y} |\cO(x+iy)|\exp[-(\log(1+x^2+y^2))^2]\,,
\ee
which contains all polynomials in $z=x+iy$ and which also lie in the Hilbert space $\cH_S$ 
(\ref{hilbertdef}). For comparison with the drift criterion we also consider the weight 
functions $\sigma_{d,\alpha}(x,y)$ (\ref{sigmad}) with $\alpha_0=0.1$, 
$\alpha_1=0.5$,$\alpha_2=1.0$. Note that all functions $\sigma_{d,\alpha}$ grow like $\exp(\alpha 
|z|^3)$, much more strongly than $\sigma$.

\begin{center}
\begin{figure}
\includegraphics[width= 0.45\columnwidth] {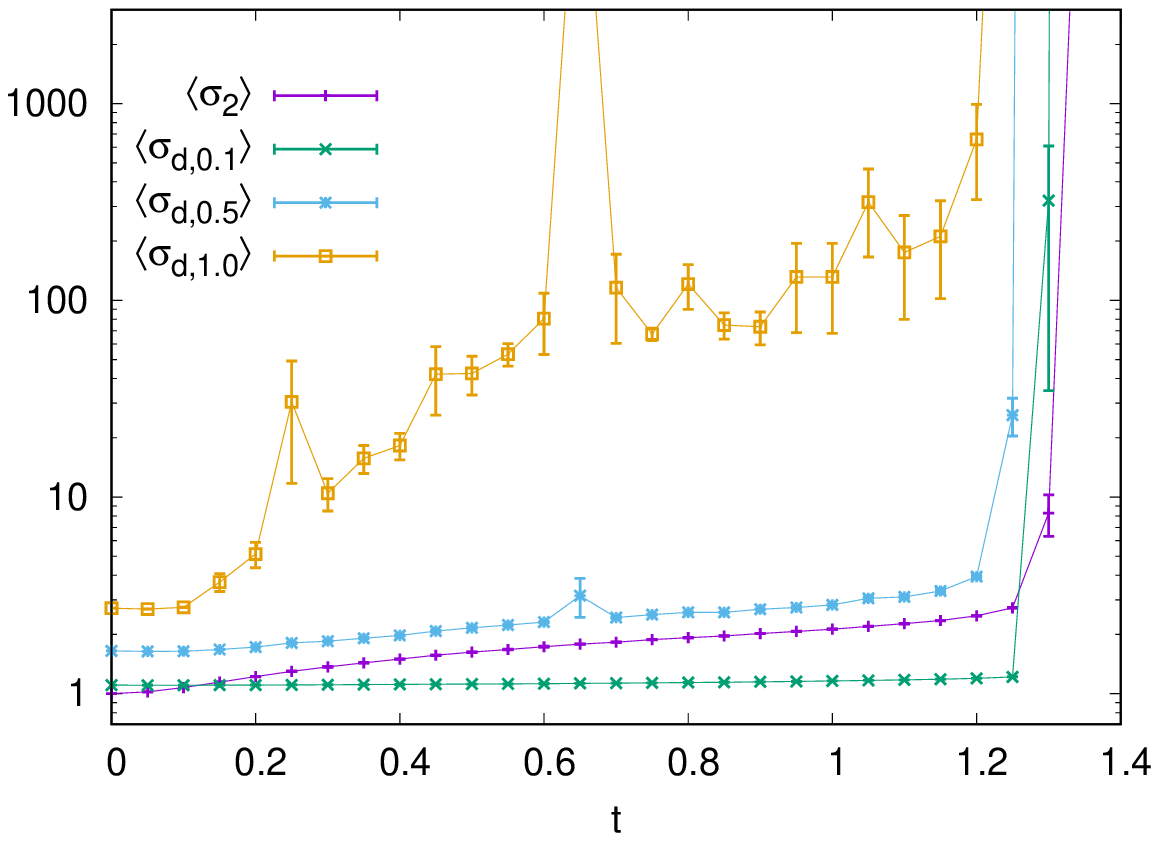}
\includegraphics[width= 0.45\columnwidth] {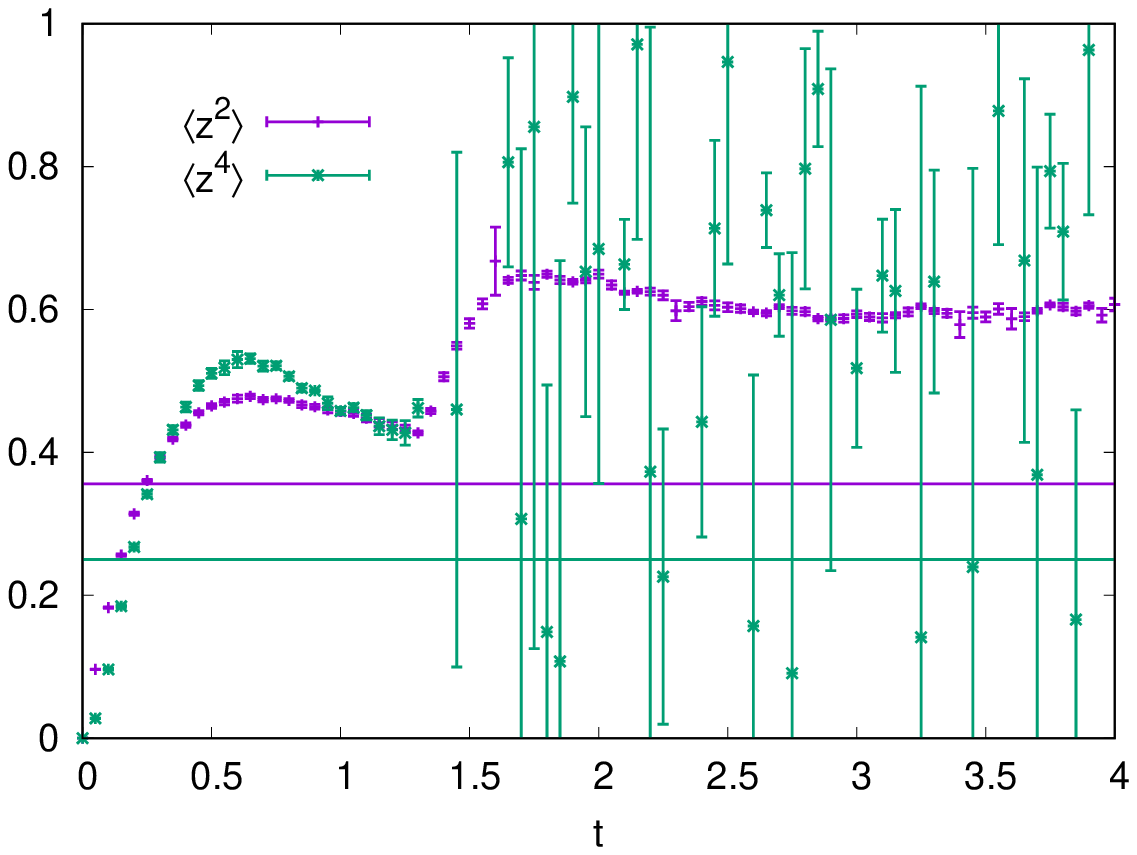}
\vglue0cm
\caption{$\bra \sigma_2\ket_t$, $\bra \sigma_\alpha\ket_t$, $\alpha=0.1,\;0.5,\; 1.0$
 (left) and $\bra z^2 \ket_t$, $\bra z^4 \ket_t$ (right) vs. Langevin time $t$, all averaged over 500000 
trajectories starting at the origin; parameters $\lambda=1$, $m^2=0.1$, $h=1.0i$. The 
horizontal lines give the `correct' equilibrium values.}
\label{fpe}
\end{figure}
\end{center}

In Fig.~\ref{fpe} we show the expectation values $\bra \sigma_2\ket_t$ and $\bra 
\sigma_{d,\alpha}\ket_t$ ($\alpha=0.1,\;0.5,\; 1.0$) under the CL process for short Langevin times 
$t$, for the parameters $\lambda=1$, $m^2=0.1$, $h_I=1.0$. We see a sharp increase of both 
quantities starting above $t=1.2$; $\bra\sigma_{d,1.0}\ket_t$ takes off a little earlier than 
the other expectation values; this is not surprising considering the stronger growth for 
$\alpha=1.0$. So for these parameters, Assumption $A$ is violated with all choices of the weight function.

We also conclude from Fig.~\ref{fpe} that for the parameters chosen and $t<1.2$ there are no visible boundary terms. This means that via integration by parts
\be
\int P(x,y;t)\cO(x+iy) dx\,dy= \int P(x,y;0) \cO(x+iy;t) dx\,dy\,.
\label{nobt}
 \ee
Notice that increasing $t$ up to $1.2$ the data tend towards the correct equilibrium values, but before they can reach them, boundary terms appear and drive the results away from the correct ones (for $\bra z^2 \ket$) or make it impossible to determine them due to huge fluctuations (for $\bra z^4 \ket$). In other models \cite{boundaryterms,boundaryterms2} it was found that for suitable choice of parameters, there is a `plateau' in $t$,
i.~e. an interval in $t$ in which the CL results were consistent with the correct ones, before the boundary terms appeared. For the model at hand, this situation also occurs for smaller values of $h_I$.

The distribution appears for $t<1.2$ to decay faster than $1/\sigma_2$ and 
$1/\sigma_{d,\alpha}$, for larger $t$ the decay becomes slower, probably power-like, leading to boundary terms and failure of CL.

It is possible to understand why around $t=1.2$ the character of the CL process changes
 and a skirt begins to show up: since we have no noise in the imaginary ($y$) direction,
 motion in this direction cannot be faster than that determined by the deterministic equation 
 \be
 \dot y= K_y\,,\ \ \quad  K_y(x,y)=-m^2 y -3 \lambda x^2 y  + \lambda y^3 -h_I. 
 \label{determ}
 \ee
 The flow pattern of the drift is such that $|K_y|$ has the largest downward size for $x=0$, in fact the solution of (\ref{determ} will reach $-\infty$ after a finite time $t_c$. Since we are starting the process at the origin, for a time $t<t_c$, in the 
 presence of noise in the $x$ direction, no value lower than $y_0(t)$ can be reached; so $P(x,y;t)$ is supported in a strip $y_0(t)<y<0$ and it is well localized in $x$, so no skirt can arise. $t_c$ is determined from (\ref{determ}) by
 \be
 t_c= \int_0^{-\infty} \frac{dy'}{K_y(0,y')}\,;
 \ee
 for $h_I=1$ this gives
 \be
 t_c=1.25125.
 \ee

\begin{figure}
\includegraphics[width= 0.45\columnwidth] {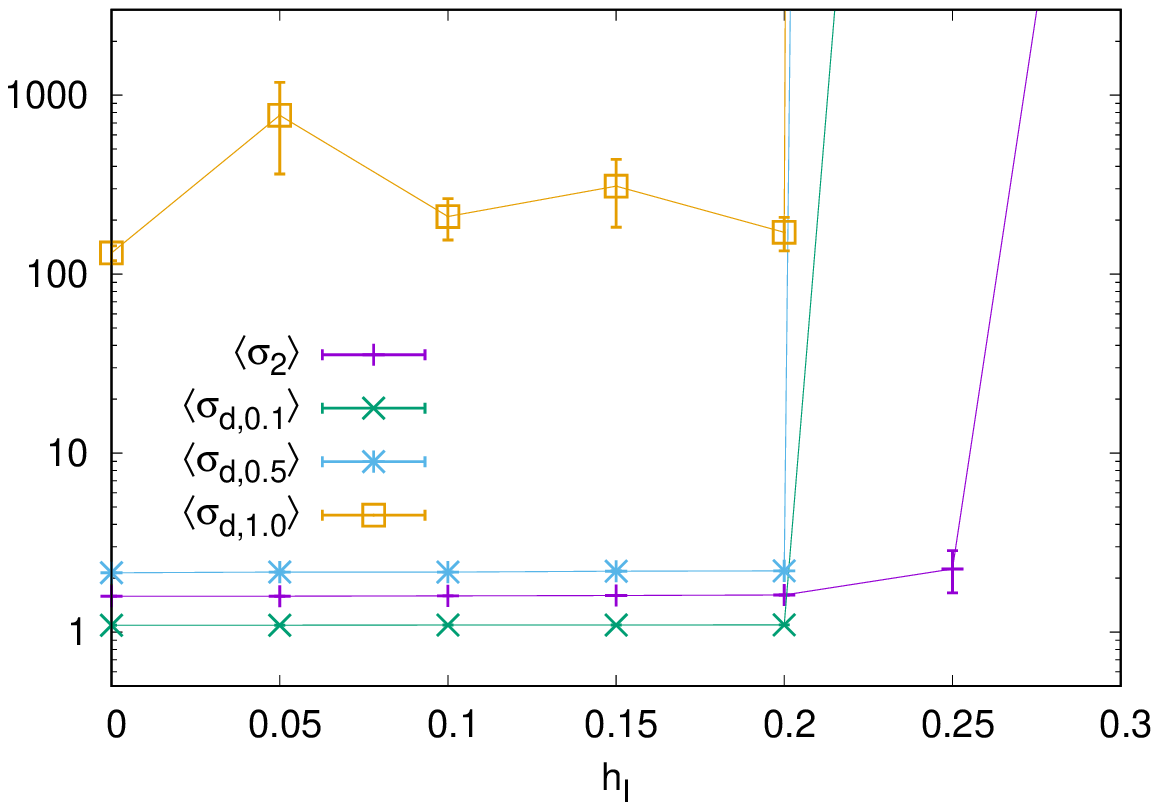}
\includegraphics[width= 0.45\columnwidth] {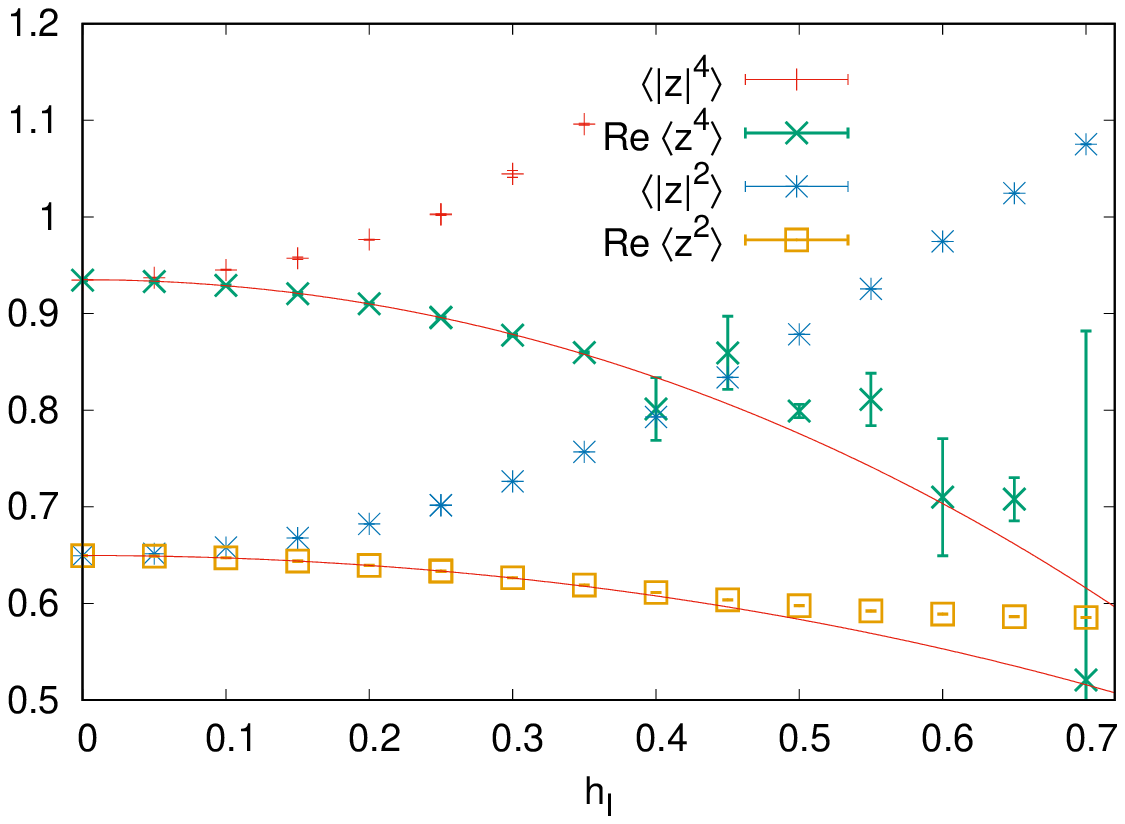}
\caption{Equilibrium CL results. Left: $\bra \sigma_2\ket_t$ and $\bra 
\sigma_\alpha\ket_t$, $\alpha=0.1,\;0.5,\; 1.0$ vs. $h_I=\Im\ h$, at $\Re\ h=0$. Right: $\textrm{Re} \bra z^2\ket$ and $\textrm{Re} \bra z^4\ket$ as well as $\bra |z|^2\ket$ and $\bra |z|^4\ket$ vs. $h_I=\Im\ h$, at $\Re\ h=0$; red lines: exact 
results.}

\label{control}
\end{figure}

In Fig.~\ref{control} (left panel) we present equilibrium values for $\lambda=1$, $m^2=0.1$, 
varying $h_I$ from $0$ to $1$. We show expectation values of $\sigma_2$, as well as 
$\sigma_{d,\alpha}$ ($\alpha=0.1,\;0.5,\; 1.0$); in the right panel we show $\Re \bra z^2 \ket $ and 
$\Re \bra z^4 \ket $ as well as $\bra |z|^2\ket$ and $\bra |z|^4\ket$.  We see that $\bra 
\sigma_2 \ket$ blows up at $h_I=0.25$, while $\bra \sigma_{d,\alpha}\ket $ blow up at already at $h_I=0.2$; $\bra z^4 \ket$ starts 
deviating from the exact value around $h_I=0.3$, whereas $\bra z^2\ket$ starts showing already 
considerable deviation from the exact value starting at $h_I=0.4$. 
At $h_I=0.3$ the observable $\bra z^4\ket $ starts to show increasing errors, while the control
variables show huge values $\log (\bra \sigma_2\ket ) \simeq O(10-100)$.

We should remark it does not make sense  to expect correctness for low powers and failure for the 
higher ones, because that would mean failure of the `consistency conditions' linking different
powers (see \cite{etiol}). This again shows that it is necessary to work in a space of observables 
invariant under $L_c$ and $\exp(tL_c)$.

We should also note that with our choice of purely imaginary $h$,
the exact values of $\bra z^n\ket$ for $n$ even are purely real, 
whereas for $n$ odd they are purely imaginary. As for the CL values, due to the symmetry 
$x\to-x$ of the drift force $K$, this is also true as long as we have convergence.

We conclude from Fig.~\ref{control} that for $\lambda=1$, $h_I\le 0.25$ there are no visible boundary 
terms; the values of $\Re \bra z^n\ket$ agree within the errors with the 
exact ones (the same is true for the imaginary parts) and $L^T$ as well as $L_c$ have no spectrum in the right hand plane.

It should be noted that the first Lee Yang zero for our choice $\lambda=1, m^2=0.1$ is 
at $h_1\approx 2.52i$, so the blowup of $\bra\sigma\ket$ both for finite $t$ 
(Fig.~\ref{fpe}) and for the equilibrium in Fig.~\ref{control} happens for $|h_I|$ much 
smaller than $|h_1|$. This is because the form of Assumption $A$ we used is sufficient 
but not actually necessary to rule out ``unstable modes''. In the next subsection it is 
shown by direct numerics that spectrum in the right half plane only appears for 
$|h_I|>|h_1|$.

It is also noteworthy that apparently it does not make much difference whether one chooses $\sigma_2$ 
or one of the $\sigma_{d,\alpha}$ as control variables as long as $\alpha$ is not too large: the blowup 
happens pretty much in the same place for the different $\alpha$ values, both in $t$ and in $h$.

But the simulations, together with the results of the next subsection, also make manifest that spectrum 
in the left hand plane alone does not guarantee correctness because it does not rule out boundary 
terms. The massive failure of expectation values $\bra z^2\ket$ and $\bra z^4\ket$ long before the 
first Lee-Yang zero shows this clearly. We also look at boundary terms for the observables $z$ and 
$z^2$; as explained in \cite{boundaryterms,boundaryterms2}), they are obtained as
\be
B_k(\cO)=\lim_{C\to\infty}\int_{|z|^2\le C} P(x,y;t=\infty)L_c^k \cO(z)\,\quad k=1,2,3,\ldots,
\quad \cO=z\; {\rm and}\; z^2
\ee

\begin{figure} 	
\includegraphics[width= 0.45\columnwidth] {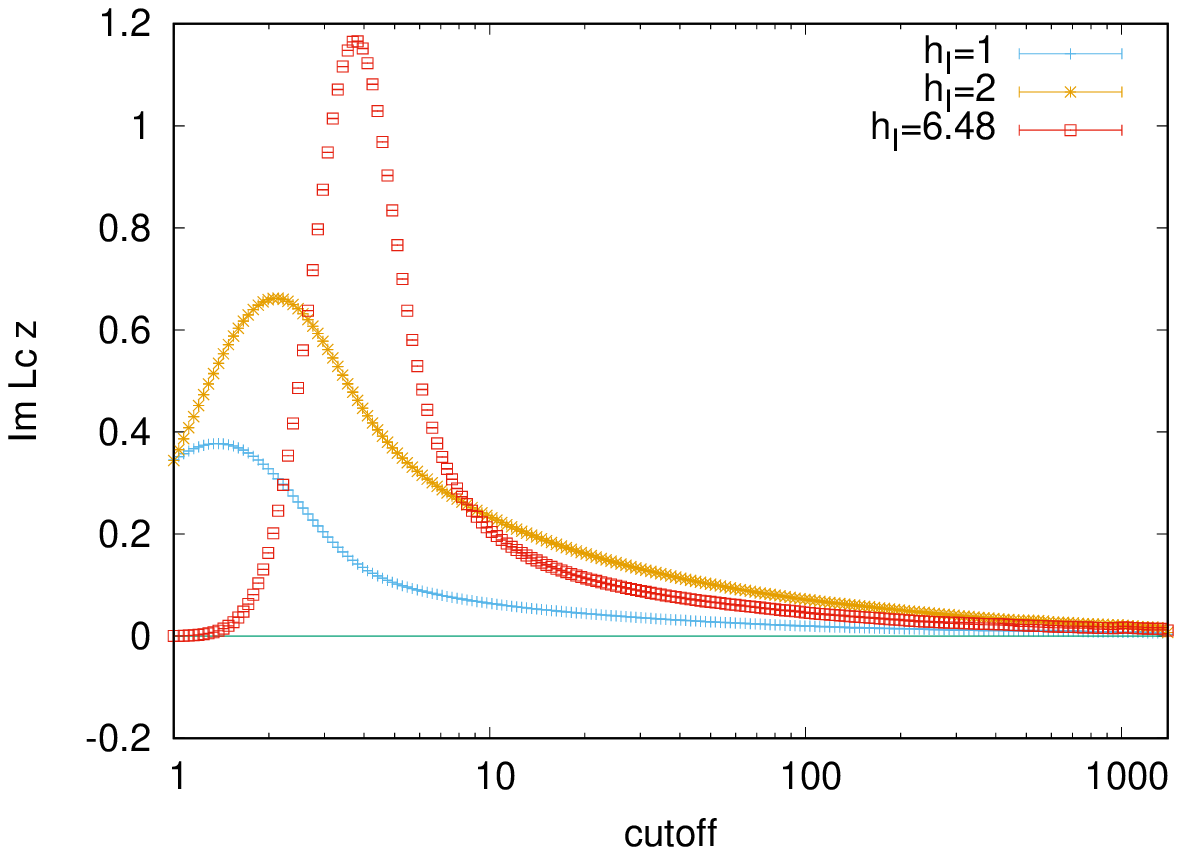} 
\includegraphics[width= 0.45\columnwidth] {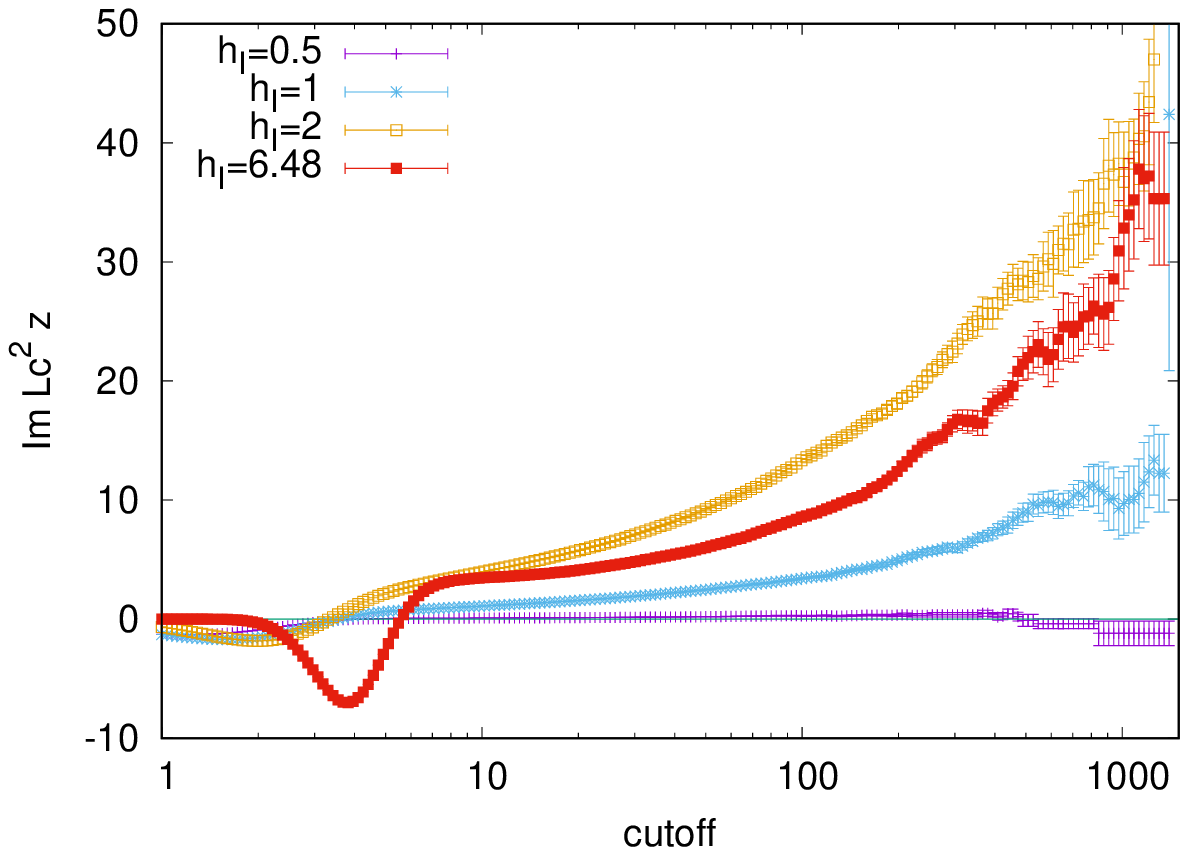} 
\includegraphics[width= 0.45\columnwidth] {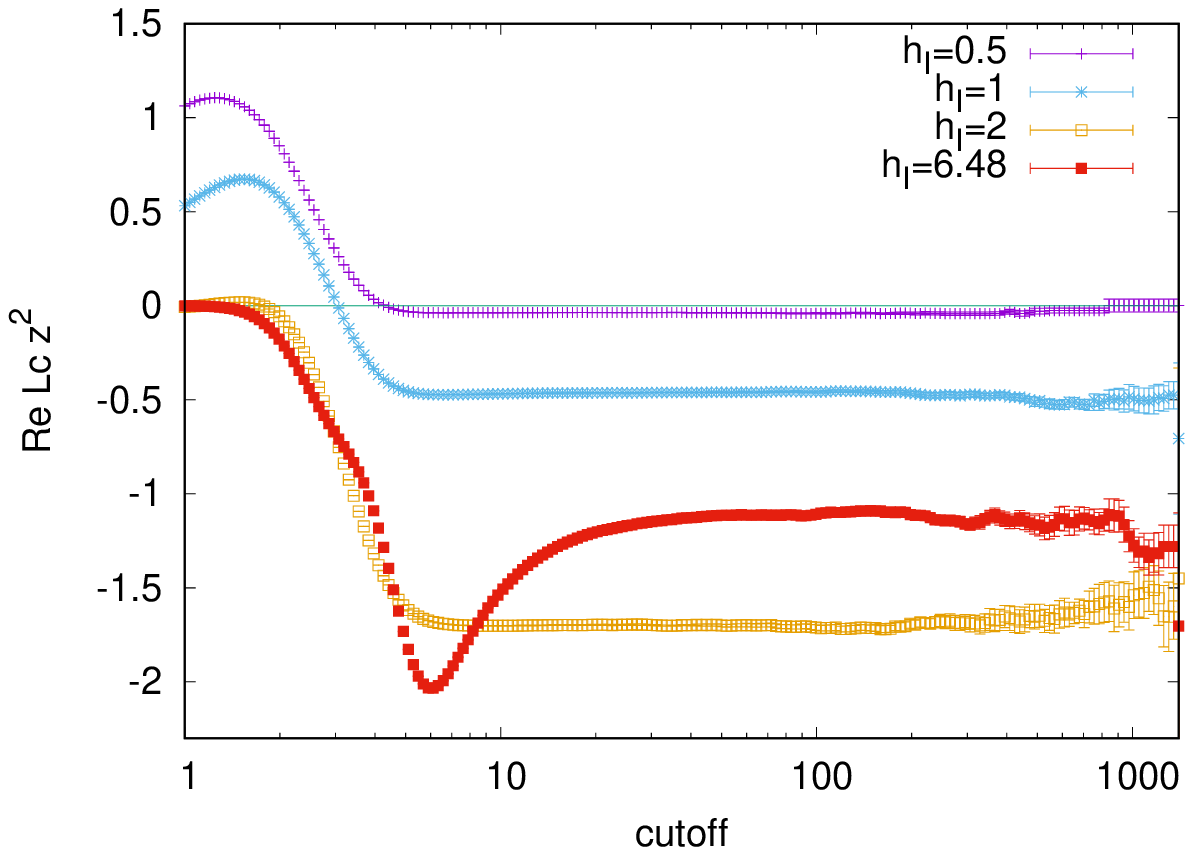} 
\caption{ 
The boundary terms in the quartic model (\ref{salcedo}) for $m^2=0.1$, $\lambda=1$ and 
$\textrm{Re}~h=0$: the imaginary parts of the first and second boundary terms of the observable $z$ and 
the real part of the first boundary term of the observable $z^2$ is shown as a function of the cutoff 
$C=|z|^2$, for various $ h_I=\textrm{Im}~h$ values, as indicated.
}
\label{toyboundary}
\end{figure}

In Fig.~\ref{toyboundary} we show the plot of the boundary terms of these observables. As one observes, 
the first boundary term of $z$ seems to be consistent with zero (in the infinite cutoff limit) for all 
magnetic fields, however the second boundary term is nonzero above $h_I>0.5 $. For the 
observable $z^2$, already the first boundary term shows nonzero values above $h_I>0.5$. This 
confirms the assessment made above using the control variables $\sigma_2$ and $\sigma_{d,\alpha}$.

%% file: 3bsalcedo.tex

\subsection{Direct determination of the spectrum}

In this section we investigate the spectrum of the $L_c$ operator for the model (\ref{salcedo}), 
given by
\bea
L_c = \partial_z^2 + K(z) \partial_z = \partial_z^2  - ( m^2 z + \lambda z^3 + h) \partial_z.  
\eea

For the numerical investigation we used several bases $e^{(a)}_i$:
\bea
e^{(1)}_n= z^n, \ \qquad
e^{(2)}_n = z^n e^{-z^2} , \ \qquad
e^{(3)}_n = z^n e^{-S(z)/2}, \ \qquad
e^{(4)}_n = \Phi_n(z), \ \qquad
\eea
where $ e^{(1)}_n $ is the monomial basis, and $\Phi_n(z)$ are the eigenfunctions of the 
Hamiltonian of the corresponding harmonic oscillator.
We truncate these bases using the first $N$ basis vectors, and calculate the 
spectrum of
the resulting $N \times N$ (in general complex) matrix using the QR algorithm
with explicit shifts \cite{golub2013matrix}.
The numerical diagonalization requires the usage of high precision
numbers, e.g. at $N=1024$ we use floating point numbers
with a mantissa of 1024 bits to avoid the appearance of spurious eigenvalues
due to precision loss.

 To transform $L_c$ into the bases given above one writes e.g. 
 \bea
 \cO(z,t) = \sum \alpha_n(t) z^n e^{-Az^2 },
 \eea
 where $A=0$ gives the monomial base $e^{(1)}$ and $A=1$ gives the base
 $ e^{(2)} $.
One than calculates 
 \bea
 (\partial_z+K(z)) \partial_z \cO  = 
 \sum_n \alpha_n & \left[  n (n-1) z^{n-2} -4 A z^n n - 2 A z^n + 4A^2 z^{n+2}
   -m^2 n z^n -\lambda n z^{n+2}  - h n z^{n-1} \right. &
 \\ \nonumber
  &  \left. + 2A m^2 z^{n+2} + 2 A \lambda z^{n+4} + 2 A h z^{n+1} \right] 
 e^{-Az^2} &
 \eea
 Thus $L_c$ in this basis is given by


 \bea
 (L_c)_{nk} & = & (n+2) (n+1) \delta_{n+2,k}  -(4 A n + n m^2+2A)  \delta_{n,k}
 +  (4 A^2 - (n-2) \lambda +2A m^2   ) \delta_{n-2,k}
 \\ \nonumber
 && +  2 A \lambda   \delta_{n-4,k}
  +   2 A h   \delta_{n-1,k}
   + h (n+1) \delta_{n+1,k}.
 \eea
 
\begin{center}
\begin{figure}
  \includegraphics[width= 0.45\columnwidth] {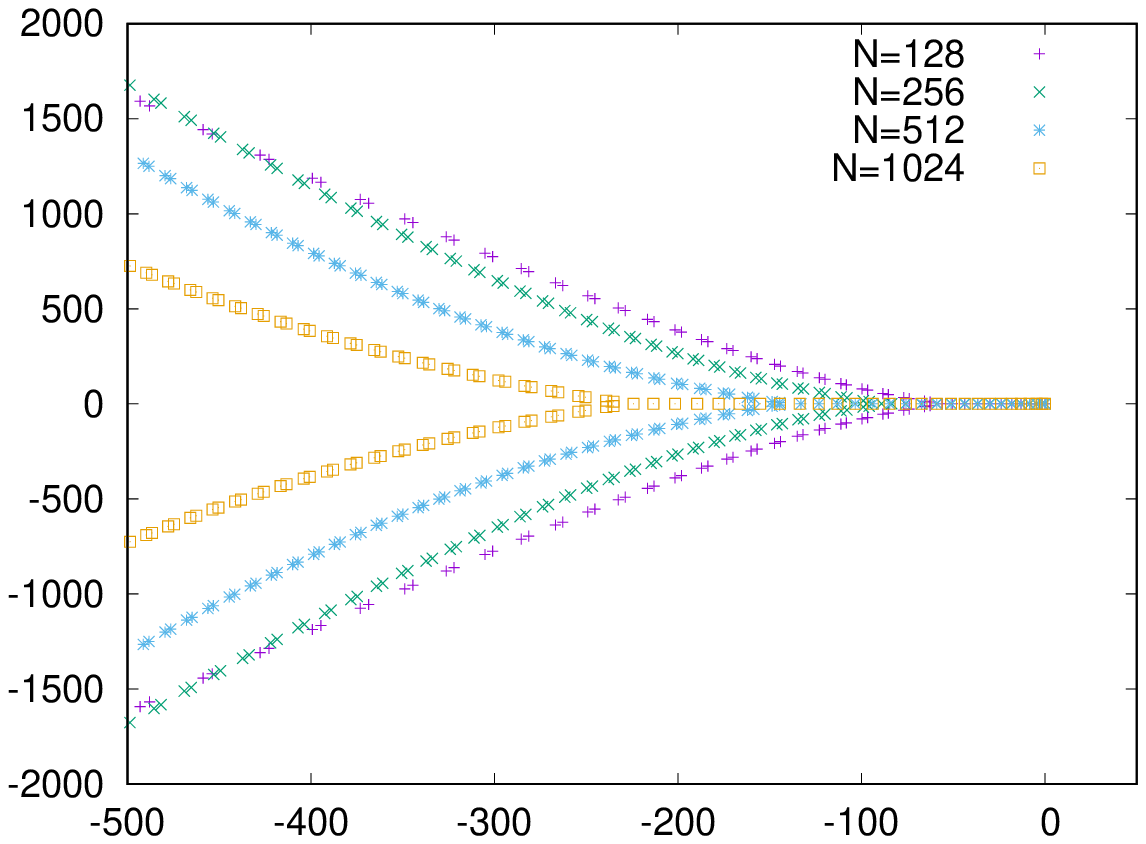}
  \includegraphics[width= 0.45\columnwidth] {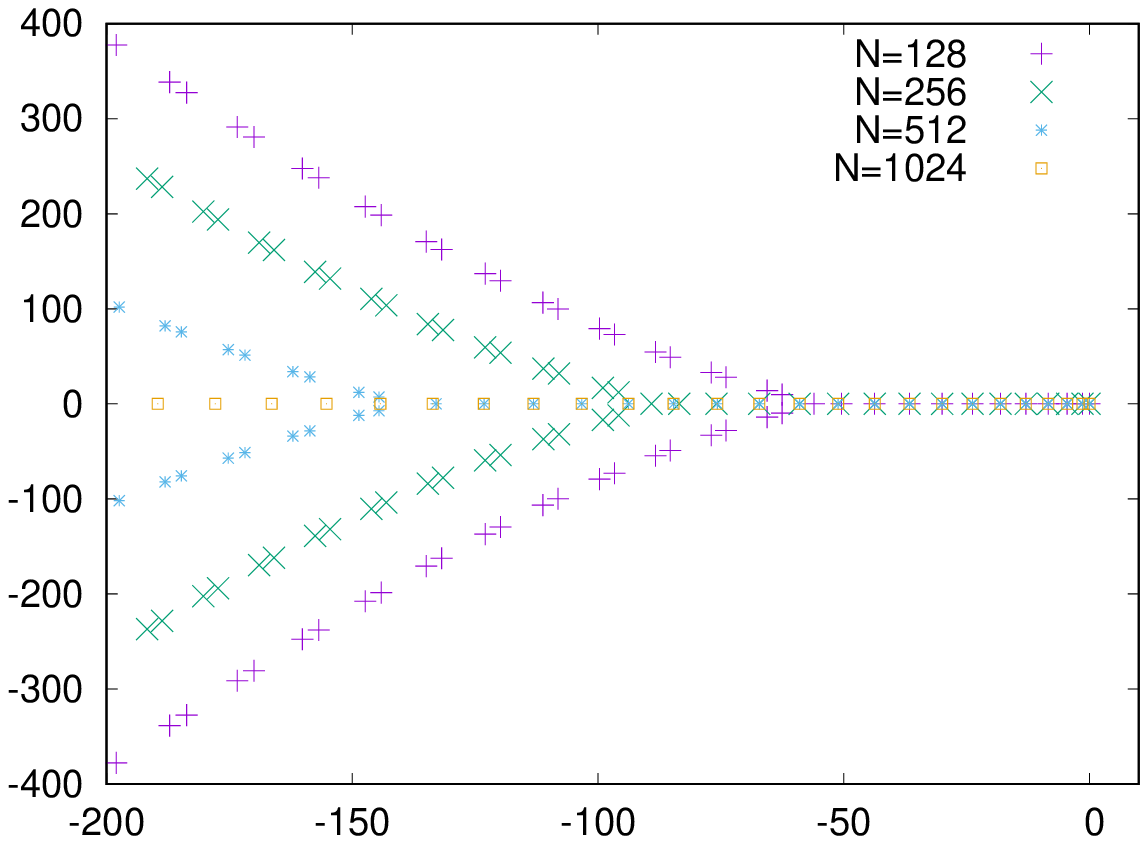}
\caption{
  The spectrum of the $L_c$ operator at $\lambda =1,\ m^2=0.1,\ h=0$.
  On the right panel, a part of the spectrum closer to zero is shown.
}
\label{fig:zerospec}
\end{figure}
\end{center}
First we investigated the convergence of the spectrum as the truncation
is improved. At $ h=0$, the spectrum of $L_c$ is known to
consist of non-positive real eigenvalues. 
In Fig.~\ref{fig:zerospec} we show the spectrum of $L_c$ for zero magnetic field for 
different truncations in the $ e^{(2)} $ basis. We observe that non-real eigenvalues 
appear for the truncated matrices, however, as the truncation is improved, more and 
more eigenvalues appear on the real axis, and the converged eigenvalues are all real 
(and non-positive). In principle the spectrum of the $L_c$ operator in other bases 
should converge to the same eigenvalues, provided the bases are related by a bounded 
linear map with a bounded inverse (for more general basis changes this may fail, see 
Appendix A). The convergence rate (with increasing truncation), however, may be basis 
dependent even then. From the bases mentioned above, $ e^{(2)}$ shows by far the fastest 
convergence rate.

\begin{center}
\begin{figure}
  \includegraphics[width= 0.45\columnwidth] {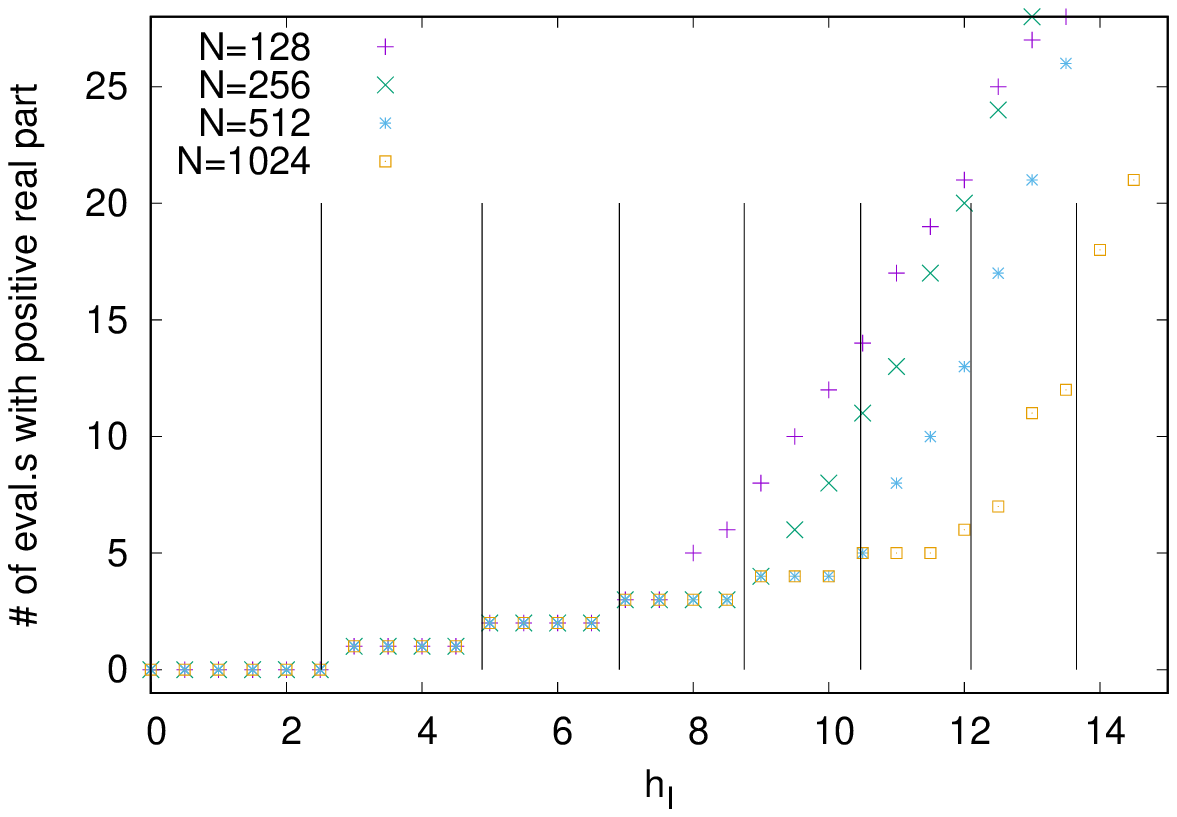}
  \includegraphics[width= 0.45\columnwidth] {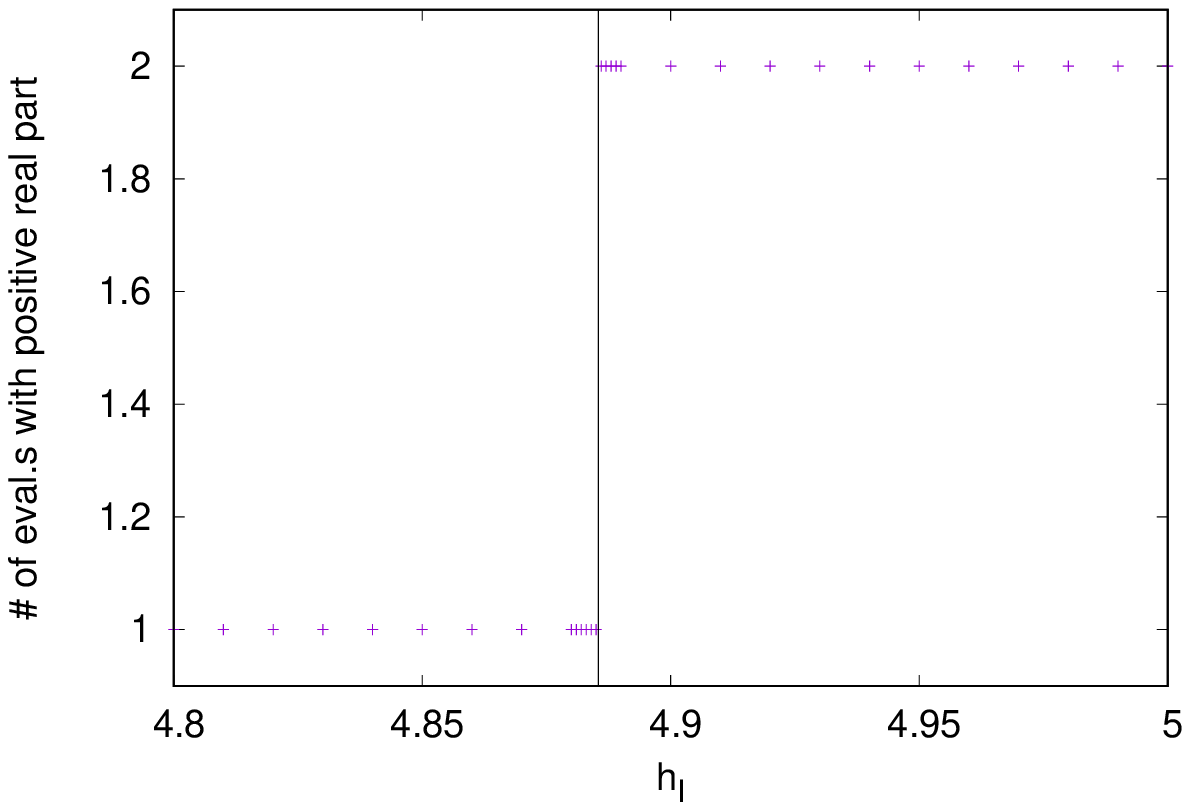}
\caption{
  The number of eigenvalues of $L_c$ with positive real part as a function of the imaginary part of magnetic field $h$, at $ \textrm{Re}\ h =0 $,
  for different truncations in the basis $e^{(2)}$.
  The Lee-Yang zeroes of the theory are indicated by vertical lines.
  Right: zoom in around the second Lee-Yang zero.
  The parameters used are $ \lambda=1, \ \ m^2=0.1 $.
}
\label{fig:Nleeyang}
\end{figure}
\end{center}

Next we investigate the number of positive real-part eigenvalues (which make the $L_c$ 
evolution of some observables unstable) as a function of the magnetic field $h$. At zero real 
part of the magnetic field, as $\textrm{Im}\ h$ is increased, eigenvalues appear with positive 
real parts. As observed in Fig.~\ref{fig:Nleeyang}, the number of such eigenvalues increases by 
one precisely at the Lee-Yang zeroes of the theory. For higher magnetic field magnitudes, 
higher truncation of the $L_c$ operator has to be used for convergence, as observed in 
Fig.~\ref{fig:Nleeyang}.

\begin{center}
  \begin{figure}
    \includegraphics[width= 0.45\columnwidth] {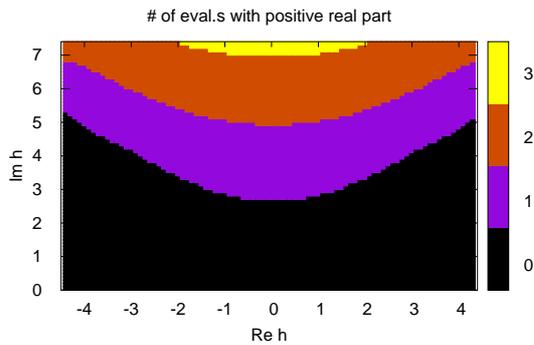}

\caption{
  The number of eigenvalues of $L_c$ with positive real part
  on the complex plane of the magnetic field $h$,
  using the truncation $N=512$, and the parameters
  $ \lambda=1, \ \ m^2=0.1 $.
}
\label{fig:evcount}
\end{figure}
\end{center}

In Fig.~\ref{fig:evcount} we show the number of eigenvalues with positive real part as a
function of the complex $h$ parameter. Note that by the Lee-Yang theorem, which applies 
here \cite{leeyang,griffiths.simon}, zeroes only occur for purely imaginary $h$.

%% file: 4xymodel.tex
\section{The XY model}
\label{xy}

In this section we test the proposed control variables of the correctness criterion for the three dimensional XY model defined by the action
\bear
S = -\beta \sum_x\sum_{\nu=0}^2  \cos \left( \phi_x-\phi_{x+\hat\nu} -i \mu \delta_{\nu,0} \right),
\eear
where $x$ represents the space-time coordinate on a 2+1 dimensional cubic lattice, $ x+\hat \nu$ 
is the neighboring lattice point of $x$ in the direction $\nu$, and $\mu$ is the chemical 
potential. At $\mu>0$ the action is in general complex, resulting in a sign problem hindering 
Monte Carlo simulations of the theory. This model has been previously investigated using the CLE 
in \cite{Aarts:2010aq}, and its boundary terms in \cite{boundaryterms2}. The sign problem of 
this model can be solved using the worldline formulation \cite{Banerjee:2010kc}.
\begin{center}
\begin{figure}
  \includegraphics[width= 0.75\columnwidth] {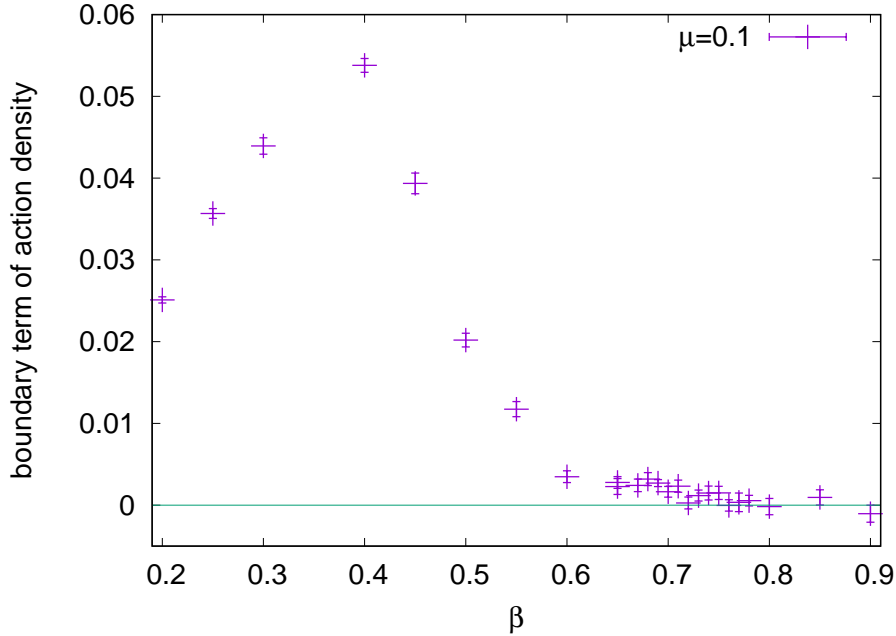}
\caption{
The boundary terms of the action density in the XY model as a function of $\beta$.  
}
\label{fig:xyactionboundary}
\end{figure}
\end{center}
To investigate the boundary terms in the XY model, we first defined the norm $ N_{IM}=\textrm{max}_x (\textrm{Im~} \phi_x)^2$, where the field 
configurations satisfying $ N_{IM}< C $ for some real $C$ enclose the real
manifold in a bounded domain.
We than investigate the observable
\bea
 \theta( C -N_{IM}  ) \frac{1}{N_s^2 N_t } L_c S[\phi_x]
\eea
as a function of the cutoff $C$. The limiting value
for infinite cutoff gives the value of the boundary term \cite{boundaryterms2}. In practice the fluctuations of the observable increase for large $C$, so one reads of the value of the 
boundary term by e.g. fitting a constant for 
large enough $C$ values. In this case we simply take the value at the cutoff $C=8$, which is in the asymptotic region for all $\beta$ values for the parameters considered here ($N_s=N_t=8, \mu=0.1$).
It was observed in \cite{Aarts:2010aq} that the CLE solution gives within errorbars correct values in the high $\beta$ phase of the theory, and incorrect results for small $\beta$. 
In Fig.~\ref{fig:xyactionboundary} we show the boundary terms of the action density as a funtcion of $\beta$, confirming this behavior.

\begin{center}
\begin{figure}
  \includegraphics[width= 0.75\columnwidth] {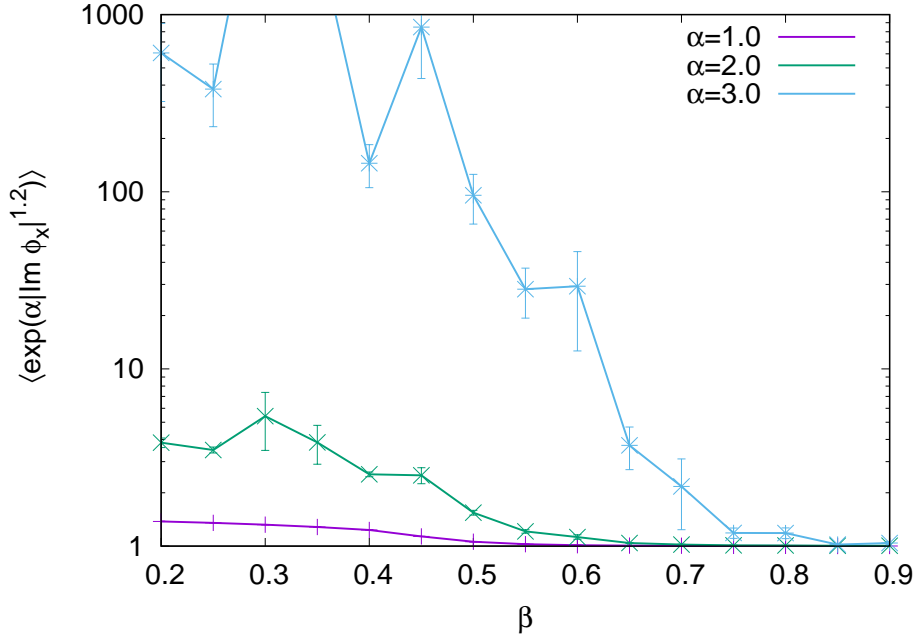}
\caption{
The ensemble average of $\sigma_{XY}$ defined in eq. (\ref{eq:sigmaxy}) as a function of $\beta$, for various $\alpha$ values, at $\mu=0.1$, measured
on an $8^3$ lattice. We used the exponent $\gamma=1.2$ as indicated.
}
\label{fig:sigmaxy}
\end{figure}
\end{center}

We introduce the weight function 
\bea \label{eq:sigmaxy}
 \sigma_{XY} (\textrm{Re}~\phi_x,\textrm{Im}~\phi_x) = \frac{1}{V } \sum_x \exp( \alpha |\textrm{Im} \phi_x |^\gamma )
\eea
for the XY model, depending on all of the complexified $ \phi $ variables on the lattice.
As discussed in the previous sections, this weight function is then used to ascertain
whether the probability distribution of the complexified process decays fast enough at infinity.
Note that $\sigma_{XY}$ has two parameters $\alpha$ and $\gamma$. The precise value of the $\gamma$ parameter is unimportant: as long 
as it is slightly above 1, the observable will be able to signal whether a faster than exponential decay in the probability
distribution of $\phi_x$ is present. 
In Fig.~\ref{fig:sigmaxy} we show the ensemble average of $ \sigma_{XY}$ at various $\alpha$ values and at $\mu=0.1$.
One observes a behavior consistent with the boundary terms: at small $\beta$ large values and large fluctuations in the 
observable signal that the observable gets most of its contribution from the tails of the distribution, and thus one should 
expect incorrect results. At large $\beta$ the values remain small, signalling fast decay and results consistent with the correct ones. 
Note that a certain experimentation with the parameter $\alpha$ is needed here, using too small(large) $\alpha$ would mean that
the observable is always small(large), however it seems there is a window of usable $\alpha$'s which correctly signal 
the behavior of the theory.

%% file: 5discussion_rev.tex
\section{Discussion and open problems}
\label{limitations}

A remaining question is if our Assumption $A$ guarantees correctness of CL. What we showed here 
is that they imply that the spectrum of $L_c$ and $L_c^T$ lies in the left half of $\C$ and there 
are no boundary terms. But this is not enough to guarantee correctness: already in \cite{trust} 
it was pointed out that correctness is only guaranteed if also $0$ is a {\em simple} (i.~e. 
nondegenerate) eigenvalue of $L_c^T$.

What we can say is the following: 

\begin{itemize} 
\item 
Absence of boundary terms in the equilibrium measure of CL ensures that the ``convergence conditions''
(CC) \cite{etiol} and the Schwinger-Dyson equations (SDE) are satisfied. This remains true also in
the presence of a kernel. 

\item 
As shown in \cite{salcse}, the SDE imply that the expectation values with the equilibrium measure are 
given by a complex linear combination of the integrations over inequivalent integration cycles. 
If there are several inequivalent integration cycles, each of them will represent a zero mode of 
$L_c^T$ (here we have to consider $L_c^T$ as an operator acting on a space of linear functionals on 
the space of observables). Integration cycles connect different zeroes of $\rho(z)=\exp(-S(z)$, 
which may be finite or infinite, or they wind around compact directions of the configuration 
space.
The real (physical) integration cycle is not always reproduced by CL, but the 
introduction of a kernel may remedy this. 

\item 
In some cases the existence of inequivalent integration cycles is also accompanied by 
nonergodicity of the CL process, i.e. the existence of different equilibrium distributions 
depending on the starting point. This will mean that the real Fokker-Planck operator $L^T$ has 
more that one zero mode. But there are also examples where the CL process is ergodic, yet 
the eigenvalue $0$ of $L_c^T$ is degenerate (see for instance the example in \cite{salcse}).

\end{itemize}

Examples of the ``mixing'' of several integration cycles compromising the correctness of the CL 
simulations are plentiful, see for instance \cite{aarts_zeroes,salcse}, as well as in 
Appendix B of \cite{alvestad}, where for certain kernels the spectrum
of $L_c$ is no longer on the left hand side of the complex plane 
(which is signalled correctly by the criterion developed in this paper), 
the boundary terms seem to vanish, and in fact the results can be 
expressed as a complex linear combination of the integration cycles \cite{salcse}.

The special case of 
non-ergodicity occurs in simple models with zeroes in the complex density $\rho$ in 
\cite{aarts_zeroes}. If $\rho$ has a finite zero on the original real integration cycle,
there are typically at least two inequivalent integration cycles, starting at that zero and going
to infinity in different directions. More zeroes lead to supplementary cycles
connecting 2 zeroes. Such cycles also occur or in the case of compact models,
connecting two finite zeroes. In fact, here it is easy to see that the eigenvalue $0$ of $L_c^T$ is degenerate: we may
multiply $\psi_0=\exp(-S)$ by the characteristic function of an interval between two zeroes (one
of which may be at infinity), thereby producing a new eigenfunction of $L_c^T$ with eigenvalue
$0$.

But ergodicity may also fail in simple quartic models without finite zeroes, 
for instance for
\be 
S=\frac{\lambda}{4}(z^2-(a+ib)^2)^2\,,\quad \lambda=2.\,,a=3.\,,b=1.\,, 
\ee
where numerics strongly suggests that there are two different equilibrium 
distributions\cite{unpub}. Another example of apparent nonergodicity is found in Appendix B of \cite{alvestad}, in that case involving CL with a constant kernel.

In a non-ergodic situation, in particular in the presence of zeroes of $\rho$ , it may depend on 
the starting point of CL which integration cycle or which linear combination of cycles is 
represented. For lattice models it is of course quite difficult to determine all the possible 
integration cycles as well as the linear combination of them representing the original problem.

To summarize, we have located the main problems of the Complex Langevin method:
First, insufficient decay of the probability distribution generated by the process, which leads to boundary terms and spoils the averages. Second,
degeneracy of the zero mode of $L_c^T$,  which is related to inequivalent integration cycles of the theory (this includes ergodicity problems).
The first problem has been thoroughly studied both in simple models and in lattice simulations of realistic models. It can be tested for using an on-line measurement, sometimes even correction of the CL results can be performed \cite{boundaryterms,boundaryterms2}. In this paper we have proposed 
some diagnostic observables which signal the first as well as the second problem.

These problems (especially the second one) need further investigation, probably with the introduction of (field dependent) kernels.
If a kernel has the effect of forcing the equilibrium distribution to stay close to the 
real, respectively unitary (physical) manifold, this could alleviate 
the aforementioned problems and the results will typically be correct
\cite{alvestad,Boguslavski:2022dee,Lampl:2023xpb,Alvestad:2023jgl}.

%% file: 6appendixa.tex
\appendix\section{Some subtle points concerning the spectrum of non-selfadjoint operators}
\label{spect}

Since in the 1980's the CL pioneers Klauder and Petersen \cite{klauder_petersen} lamented
about the {\it \ldots conspicuous absence of general spectral theorems \ldots } there has been a lot
of research on this issue; several textbooks have appeared, which deal with the thorny
question of the spectrum of unbounded operators which are not normal operators on a Hilbert space, see e.~g. \cite{vanneerven,engelnagel,daviesbook}. There are, however, still many open problems. We want to mention a few unpleasant facts showing that the situation is much more subtle than in the case of normal operators or finite matrices.

1. The spectrum is not always invariant under similarity transformations. A simple example
is found in \cite{daviesbook}, Example 9.3.200: consider
\be
L=\partial_x^2+b\partial_x \quad {\rm on}\; \cL^2(\R).
\ee
By a similarity transformation well-known from CL, $L$ is transformed into $-H$
\be
-H=\exp(bx/2) L \exp(-bx/2)= \partial_x^2+\frac{b^2}{4}.
\ee
The spectra are easily seen, using Fourier transformation, to be
\bea
{\rm spec}(L)&=\{\lambda\in \C\,|\,\lambda=-p^2+ibp, p\in \R\}\notag\\
{\rm spec}(-H)&=\{\lambda\in \R\,|\,\lambda=-p^2-b^2/4, p\in \R   \}\,,
\eea
i.e. a parabola vs. a half line.

2. The fact that the spectrum of $L$ is in the left half plane does not preclude growth of
the semigroup $\exp(tL)$, even in the finite dimensional case.

The simplest example is given by $L=a^\dagger$ where $a^\dagger$ is the  creation operator of
one fermion mode. The semigroup is $\exp(ta^\dagger)=1+ta^\dagger$ which shows linear growth,
even though
\be
{\rm spec}\,(a^\dagger)=\{0\}\,,\quad {\rm spec}(\exp(ta^\dagger))=\{1\}\,.
\ee
There are examples of much stronger sub-exponential growth, given by Volterrra operators,
whose spectrum consists again just of the origin.

But even exponential growth can happen for an operator whose spectrum lies entirely in
the closed left half plane, see \cite{daviesbook}, Theorem 8.2.9, which discusses an
infinite matrix example due to Zabczyk.

3. It is not true in all generality that the spectrum of $\exp(tL)$ is given by the
exponential of the spectrum of $L$. This requires that the so-called spectral mapping
principle holds. A detailed discussion is given in \cite{vanneerven}, Ch.~2.

4. A useful condition is the following:

A closed operator $L$ is called {\em dissipative}, if
\be
\Re~(\psi,L\psi)<0
\label{diss}
\ee
for all $\psi$ in the domain of definition of $L$. There is a theorem, due to Lumer and
Phillips \cite{engelnagel}, Cor.~3.17, showing that $L$ defines a contractive semigroup,
i.e. $||\exp(tL)||<1$ $\forall t>0$, if $L$ is dissipative. The converse is also true,
i.~e. if $L$ does not satisfy (\ref{diss}), $\exp(tL)$ will not be contractive.

5. Considering our Assumption $A$, it seems a natural setting would be in the context of
Banach, rather than Hilbert spaces. In fact the book by van Neerven is written that way.

%% file: 7appendixb.tex
\section{Example of non-unique $L_c$ evolution}
\label{nonunique}

The {\em real} one-pole model in the simplest case is defined by
\be
 \rho(x)= x^2 \exp(-\beta x^2)\,.
\ee
The kernel of $\exp(tL_c)$ is given in \cite{seiler} as
\bea
\exp(tL_c)(x,y)
=&2\frac{y}{x}\exp\left(\frac{\beta}{2}(x^2-y^2)\right)\exp(2\beta t)
\sqrt{\frac{\beta}{\pi (1-e^{-4\beta t})}}\notag\\ \times
&\exp\left[-\frac{\beta (x^2+{y}^2)}{2\tanh(2\beta t)}\right]
\exp\left(\frac{\beta x y}{\exp (2\beta t)}\right)\,.
\label{lc_semigroup_full}
\eea
It is easy to see that this has the unstable eigenmode $1/y$ with eigenvalue $2\beta$ for
$L_c$, as found in \cite{seiler}. The initial value problem
\be
\partial_t \cO(z;t)= L_c \cO(z;t) \quad {\rm with}\; \cO(z;0)=\frac{1}{z}
\ee
thus has the solution
\be
\cO(z;t)=\exp(2\beta t) \frac{1}{z}\,.
\ee
This solution is valid for $z\in\C\setminus \{0\}$ and is jointly analytic in $t,z$.
As found in \cite{seiler}, the unstable mode is signaled also by a boundary term.

But as remarked in the appendix of \cite{seiler}, this solution is not unique, even
in the limit $\beta\to 0$.  A second solution for $\beta=0$ is
\be
\cO_{-1}(z;t)=\frac{1}{z}{\rm Erf}\left(\frac{z}{2\sqrt{t}}\right)\,;
\label{1/zhalf}
\ee
for nonzero $\beta$ it is
\be
\cO_{-1}(z;t)=\frac{1}{z}\exp(2\beta t) {\rm Erf}(c(\beta,t)z)
\ee
with
\be
c(\beta,t)=\sqrt{\frac{\beta \exp(-4\beta t)}{1-\exp(-4\beta t)}}
\ee
This solution is for $t>0$ holomorphic in $z\in\C$, but it has an essential singularity
at $t=0$. On closer inspection it is seen that as soon as $|\Im z|>|\Re z|$, the solution blows
up as $t\to 0$, so it does not solve the initial value problem everywhere. It solves it only in the wedge $|\Im\,z|<|\Re\,z|$. 
 
The solution (\ref{1/zhalf}) is correctly represented by the {\it} real Langevin
process on the positive or negative real half-axis. There is no boundary term and
no unstable mode.

But if we want to simulate the one-pole model on a line parallel to the real axis,
only the first solution can be used. The CL process then produces a superposition
of the positive and negative half-lines (the coefficients actually depend on the
starting point of the process), as shown in \cite{salcse}; the CL process is not
ergodic. There is an unstable mode present, CL produces an incorrect result and
there is a boundary term at the origin term signaling incorrectness \cite{seiler}.